\documentclass[twocolumn,showpacs,preprintnumbers,amsmath,amssymb,superscriptaddress,showkeys,pre]{revtex4}

\usepackage{graphicx}
\usepackage{dcolumn}
\usepackage{bm}
\usepackage{textcomp}
\usepackage[letterpaper,left=2.0cm,right=1.2cm,bottom=4.5cm,top=2cm]{geometry}

\usepackage{hyperref}
\newcommand{\eg}{{\it e.g.\/}}
\newcommand{\ie}{{\it i.e.\/}}
\newcommand{\etal}{{\it et al.\/}}
\urldef\dmrgpp\url{http://www.ornl.gov/~gz1/dmrgPlusPlus/}
\hypersetup{colorlinks=true,
        citecolor=blue        
}

\begin{document}

\title{Time Evolution with the DMRG Algorithm:\\
A Generic Implementation for Strongly Correlated Electronic Systems}

\author{G. Alvarez}
\affiliation{Computer Science \& Mathematics %
Division and Center for Nanophase Materials Sciences, Oak Ridge National Laboratory, %
 \mbox{Oak Ridge, TN 37831}, USA}

\author{Luis G.~G.~V. Dias da Silva}

\affiliation{Instituto de F\'{\i}sica, Universidade de S\~ao Paulo, C.P.
66318, S\~ao Paulo, SP, Brazil, 05315-970.}

\author{E. Ponce}
\affiliation{Electrical \& Computer Engineering and Computer  %
Science Department, %
Polytechnic University of Puerto Rico, %
\mbox{San Juan, PR 00919}, USA}

\author{E. Dagotto}
\affiliation{%
Department of Physics and Astronomy, University of Tennessee, Knoxville, Tennessee 37996, USA, and %
Materials Science and Technology Division, Oak Ridge National Laboratory, Oak Ridge, Tennessee 37831, USA}
\date{\today}

\begin{abstract}
A detailed description of the time-step-targetting time evolution method within the
 DMRG algorithm is presented.
 The focus of this publication is on the implementation of the algorithm, and
on its generic application.
 The case of one-site excitations within a Hubbard model
is analyzed as a test for the algorithm,  using open chains and two-leg ladder geometries. 
The accuracy of the procedure in the case of  the recently discussed
holon-doublon photo excitations of Mott insulators is also analyzed. Performance and parallelization issues
are discussed. In addition, the full open-source code is provided as supplementary material.
\end{abstract}

\maketitle

\section{\label{sec:intro}Introduction}


The accurate calculation of  time-dependent quantum observables in correlated
electron systems is crucial to achieve further progress in this field of research,
where the vast majority of the computational efforts in the past have mainly focused
on time-independent quantities.  This issue lies at the core of the study
of
a broad array of physical phenomena in transition metal oxides and
nanostructures such as electronic transport, optical excitations, and
nonequilibrium dynamics in general. 
Accurate studies of time-dependent
properties will advance the fields of spintronics, low dimensional correlated
systems, and possibly quantum computing as well.
For a list of recent efforts on these topics by our group,
and concomitant set of references for the benefit of the readers, see
\cite{re:heidrich-meisner10,re:diasdasilva09,re:heidrich-meisner09,
 re:heidrich-meisner09b,re:heidrich-meisner09c} and references therein.

The purpose of this paper is to present an explicit 
implementation of the time evolution
within the density matrix renormalization group (DMRG) method
\cite{re:white92,re:white93}. Knowledge of the widely discussed 
DMRG algorithm  to compute
static observables is here assumed. Readers not familiar with the method are
referred to published reviews
\cite{re:schollwock10,re:schollwock05,re:hallberg06,re:rodriguez02}, and to the
original publications \cite{re:white92,re:white93} for further information. 
Our main focus here is to provide a detailed
description of the implementation of the 
time-step targetting~\cite{re:feiguin05} algorithm, 
and the discussion of a few applications. We also provide full
open-source codes and additional documentation to use those codes.\footnote{See \dmrgpp.}

The present work builds upon considerable previous efforts by other groups. 
In particular, we will mainly follow 
Manmana \etal~in Ref.~\cite{re:manmana05}. The time-step targetting
procedure was also reviewed in Ref.~\cite{re:batrouni06}. Since it would not
be practical 
to describe in a short paragraph the considerable progress achieved 
in this field of research in recent years, the interested reader is 
encouraged to consult the
aforementioned reviews along with, \eg, Ref.~\cite{re:schollwock05b}, for
a historical account of the development of the methods used in the present publication.

Since our aim is to discuss a generic method
applicable to any Hamiltonian and lattice geometry, 
here we do not discuss or implement the Suzuki-Trotter
method~\cite{re:daley04,re:white04},
but focus instead only on the Krylov method~\cite{re:krylov31} for the time evolution,
as described in Ref.~\cite{re:manmana05}.
Because its implementation can be isolated, the Krylov method can be
applied in a generic way to most models and geometries without
changes, which is not the case for other methods,
such as the Suzuki-Trotter method.
The readers interested in the Suzuki-Trotter   method
should consult the ALPS project~\cite{re:bauer11}, where the time evolving
block decimation is implemented~\cite{re:daley04}.

Our goal is to compute observables of the form
\begin{equation}
\langle \phi_1| {\rm e}^{iHt} A_{0,\pi(0)} A_{1,\pi(1)}\cdots A_{a-1,\pi(a-1)} 
{\rm e}^{-iHt} |\phi_2\rangle.
\label{eq:generalcorrelation}
\end{equation}
%
where $|\phi_1\rangle$ and $|\phi_2\rangle$ denote generic quantum many-body
states. This category of observables is sufficiently broad to encompass most
time-dependent correlations, as represented by
a number $a$ of
\emph{local} operators $A_{0,\pi(0)}$ $A_{1,\pi(1)}$ $\cdots$
$A_{a-1,\pi(a-1)}$ acting on sites $\pi(0)$, $\pi(1)$, $\pi(a-1)$ of a finite
lattice,
where $\pi(i)$ denotes the lattice site on which the operator $A_{i,\pi(i)}$ acts,
and the extra index $i$ indicates that the operator can be different at each site.

An immediate application of this formalism and code is the study of the evolution of a system that is
brought out of equilibrium by a sudden excitation. This sudden excitation can be simulated
by the state of the system given by the vectors $|\phi_1\rangle$ and $|\phi_2\rangle$.
Depending on the problem, sometimes it
is more convenient to assume that the states remain unchanged but that
it is the Hamiltonian $H(t)$ that changes with time.
Another application entails the computation of time-dependent properties of systems in equilibrium,
such as the Green's function $G_{ij}(t)$.

The organization of this paper is the following.
Section~\ref{sec:method} explains in detail the Krylov method for time evolution
within the DMRG algorithm, focusing on its implementation.
Section~\ref{sec:onesiteexciton} applies the method to the case of one-site excitations,
showing a simple picture of the accuracy of the method.
Section~\ref{sec:holondoublon} extends to two-leg ladder geometries the 
results obtained using tight-binding chains, employing
holon-doublon excitations for the specific study.
Performance issues are studied in Sec.~\ref{sec:performance}, while
Sec.~\ref{sec:summary} summarizes our results.
The first two appendices contain derivations of exact results used in the paper.
The last appendix explains briefly the use of the code, and points to its documentation.

\section{Method and Implementation}\label{sec:method}

\subsection{Lanczos Computation of the Unitary Evolution}
To carry out the previously described 
program~\footnote{This section is inspired on handwritten notes
of Schollw\"ock's, from his talk
at IPAM, UCLA, at
\url{https://www.ipam.ucla.edu/publications/qs2009/qs2009_8384.pdf}} of
computing observables of the type given
by Eq.~(\ref{eq:generalcorrelation}),
the first goal is to calculate
$|\phi(t)\rangle\equiv\exp(iHt)|\phi\rangle$.
The Lanczos technique~\cite{re:lanczos50} provides a method 
to tridiagonalize $H$ into $V^\dagger T V$, where
$T$ is tridiagonal and $V$ is the matrix of Lanczos vectors.
If the number of those Lanczos vectors is $n_l$, and the Hilbert space for $|\phi\rangle$
has size $n$, then $T$ is a square matrix of size $n_l\times n_l$, and $V$ is, in general, a rectangular
matrix of size $n_l\times n$.

$V^\dagger T V$ cannot be used \emph{everywhere} as a substitution for $H$, without inducing
large errors.
But if we start the Lanczos procedure~\cite{re:dagotto94} with the vector $|\phi\rangle$ (instead
of using a random vector as is most frequently done), then we can use that 
substitution accurately in the
multiplication $H|\phi\rangle$.
However, this is not enough here, because we need to compute the exponential of $H$.
For this purpose, it has been shown~\cite{re:hochbruck99,re:hochbruck97,re:saad03}
that  $|\phi(t)\rangle\approx V^\dagger \exp(i T t) V |\phi\rangle$
with an accuracy that increases as time $t$ decreases for fixed $n_l$.
We will assume that we have taken $t$ small enough such that we can regard
the expression above to have the accuracy of the Lanczos
technique, which is usually high. In other words, if $t$ is small enough,
we will assume that using $V^\dagger \exp(i T t) V |\phi\rangle$ as a replacement for
 $\exp(iHt)|\phi\rangle$ is not worse than using $V^\dagger  T  V |\phi\rangle$ as a replacement for $H|\phi\rangle$.
This will be enough for our purposes, but
for details on the scaling and bounds of the errors made in each case as a function of $t$ and $n_l$, see, \eg,
Refs.~\cite{re:hochbruck99,re:hochbruck97}.

Since we started the Lanczos recursive procedure with the vector
$|\phi\rangle$, then $\sum_j V_{j',j} |\phi\rangle_j \propto \delta_{j',0}$.
Finally, we need to diagonalize $T=S^\dagger D S$ into a $n_l\times n_l$
diagonal matrix $D$
with diagonal elements  $d_{l'}$.
This last step is not computationally expensive,
since $T$ is a $n_l\times n_l$  matrix, as was noted before.

Putting it all together, we arrive to
\begin{equation}
|\phi(t)\rangle_i = \sum_{k,k',l,l',j} V^\dagger_{i,k} T^\dagger_{k,k'} S^\dagger_{k',l'}
{\rm e}^{i d_{l'} t} S_{l',l} T_{l,0} V_{0,j} |\phi\rangle_j,
\label{eq:lanczosevolution}
\end{equation}
for small times $t$, where the equal sign
	should be understood to be valid within the 
accuracy of the Lanczos technique~\cite{re:saad03}.
How to deal with larger times $t$ will be explained in section~\ref{sec:timeevolution}.

\subsection{Targetting States with the DMRG Algorithm}
It appears that now we could use 
Eq.~(\ref{eq:lanczosevolution}) to compute $|\phi_1(t)\rangle$ from
some vector $|\phi_1\rangle$, and likewise
$|\phi_2(t)\rangle$  starting from some vector $|\phi_2\rangle$. Then we would just
apply the operators $A_{0,\pi(0)}$ $A_{1,\pi(1)}$ $\cdots$ $A_{a-1,\pi(a-1)}$
to those states within a DMRG procedure,
to achieve our aim of computing Eq.~(\ref{eq:generalcorrelation}).
But
the DMRG algorithm is not immediately applicable to arbitrary states, such as
$|\phi_1(t)\rangle$, and was originally developed
to compute the ground state of the Hamiltonian instead.

This difficulty has been successfully overcome (see~\cite{re:schollwock05} and references therein)
by redefining the reduced density matrix
of the left block $\mathcal{L}$ as:
\begin{equation}
\rho^\mathcal{L}_{\alpha,\alpha'} =
\sum_{\beta\in \mathcal{R}}\sum_l \omega_l \Phi^\dagger(l)_{\alpha,\beta} \Phi(l)_{\alpha',\beta},
\label{eq:rdensitymatrix}
\end{equation}
where $\alpha$ and $\alpha'$ label states in the left block $\mathcal{L}$,
$\beta$ those of the right block $\mathcal{R}$, and
$\{\Phi(l)\}_l$ is a set of, as of yet, unspecified states of the superblock $\mathcal{L}\cup\mathcal{R}$.
The states $\Phi(l)$ are said to be \emph{targetted} by the DMRG algorithm.
Because of their inclusion in the reduced density matrix, these states will be obtained with
a precision that scales similarly to the precision of the ground state in the static formulation of the DMRG.
The relevance of the weights $\omega_l$ appearing in Eq.~(\ref{eq:rdensitymatrix}) will
be discussed in section~\ref{sec:timeevolution}.

Which are the states $\Phi(l)$ that need to be included in Eq.~(\ref{eq:rdensitymatrix}) to compute Eq.~(\ref{eq:generalcorrelation})?
$|\phi_1(t)\rangle$ and $|\phi_2(t)\rangle$ are certainly needed. Since observables that
include the ground state are ubiquitous,
the ground state of the Hamiltonian needs to be targetted as well in most cases.
But additional states
need  to be included in order to evolve to larger times, as we will now explain.

Our implementation follows the time-step targetting procedure of Ref.~\cite{re:batrouni06}.
We now introduce a small time $\tau$ such that, for all $t\le\tau$, Eq.~(\ref{eq:lanczosevolution}) is accurate
in the sense defined in, \eg, Ref.~\cite{re:hochbruck99}. We consider a set of $n_v$ times $\{t_x\}_x$,
$x=0,1,\cdots n_v-1$, such that $t_x<t_{x+1}$, $t_0=0$, and $t_{n_v-1}=\tau$.
For simplicity, we assume from now on that $|\phi_1\rangle=|\phi_2\rangle\equiv|\phi\rangle$
in Eq.~(\ref{eq:generalcorrelation}).
The state $|\phi\rangle$ is defined by the particular physics problem under investigation 
and we will consider particular
examples in section~\ref{sec:casestudies}.
States $|\phi(t_x)\rangle$ for each $0\le x<n_v$ can be obtained accurately from  $|\phi\rangle$
by using Eq.~(\ref{eq:lanczosevolution}) since $t_x\le\tau$.
To compute Eq.~(\ref{eq:generalcorrelation}) for all $t\le\tau$,
we target the $n_v$ states $|\phi(t_x)\rangle$ and the ground state $|\psi\rangle$ as well.

At this point it is instructive to consider a concrete class of states $|\phi\rangle$.
In a large class of problems these states are related to the ground state $|\psi\rangle$
of the Hamiltonian by
\begin{equation}
|\phi\rangle=B_{b-1,\pi'(b-1)} \cdots B_{1,\pi'(1)}B_{0,\pi'(0)}|\psi\rangle,
\label{eq:b}
\end{equation}
for
$b$ 
\emph{local}  operators $B_{0,\pi'(0)}$, $B_{1,\pi'(1)}$, $\cdots$, $B_{b-1,\pi'(b-1)}$ acting
on sites $\pi'(0),\pi'(1),\cdots,\pi'(b-1)$ of the superblock, where
$\pi'(i)$ denotes the lattice site on which the operator $B_{i,\pi'(i)}$ acts, as
explained below. The extra index $i$ indicates that the operators $B$ can be different on different sites.
Physical examples of the operators $B$ will be given in
section~\ref{sec:casestudies}.

\begin{figure}
\includegraphics[width=8cm,clip]{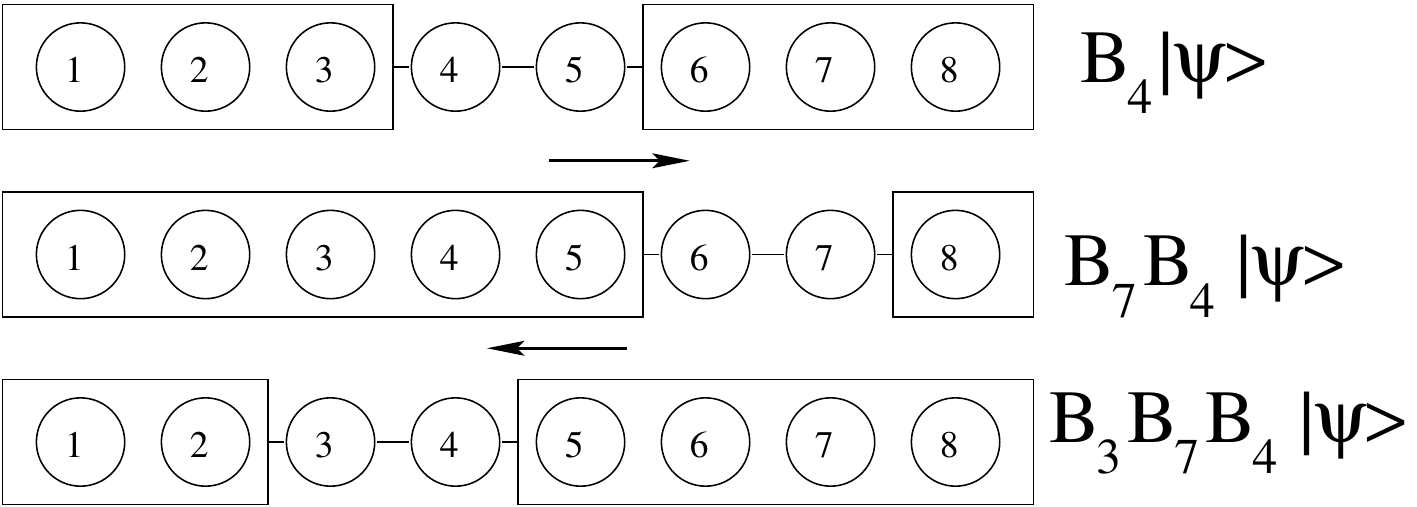}
\caption{\label{fig:DMRG_blocks} Example of a state $|\phi\rangle$ given by Eq.\ (\ref{eq:b}). The order in
which the on-site operators are applied will depend on the order in which the sites appear as ``central sites".
In the example above, $b=3$ and the operators are acting on sites 3, 4, 
and 7. In the sweeping procedure shown,
the order will be $\pi'(0)=4, \pi'(1)=7, \pi'(2)=3$, \ie,
$|\phi \rangle=B_3 B_7 B_4 |\psi \rangle$.
}
\end{figure}

The sites $\pi'(0)$, $\pi'(1)$, $\cdots$, $\pi'(b-1)$
in Eq.~(\ref{eq:b}) (as well as sites $\pi(0)$, $\pi(1)$, $\cdots$, $\pi(a-1)$ in
Eq.~(\ref{eq:generalcorrelation}))
 will be considered ordered in the way in which they appear as central
sites~\cite{re:schollwock05} for the DMRG finite algorithm,
as illustrated in Fig.\ \ref{fig:DMRG_blocks}.
At a given stage of the computational procedure, if the central site  of the DMRG algorithm is
$\pi'(0)$, then $|\phi^{\rm partial}_{0}\rangle\equiv
B_{0,\pi'(0)}|\psi\rangle$ can be obtained. Next, we proceed to the following 
site, and so on, until we
reach site $\pi'(1)$, and apply $B_{1,\pi'(1)}$, \ie, $|\phi^{\rm
partial}_{1,0}\rangle \equiv B_{1,\pi'(1)}|\phi^{\rm partial}_{0}\rangle$,
eventually reaching site $\pi'(b-1)$, to complete the computation of
$|\phi\rangle$,
given by
Eq.~(\ref{eq:b}).
Since in cases of physical interest the operators $B$ are either bosons or
fermions, a reordering is always possible due to their commutativity or
anti-commutativity, yielding at most a minus sign.
%

As the DMRG algorithm sweeps the entire lattice, the central sites change, leading to modified Hilbert spaces.
Therefore, a procedure is required to ``transport'' the states $|\phi\rangle$ from one space to another.
It is known~\cite{re:manmana05} that the transformation needed to ``transport'' these states
is the so-called \emph{wave-function transformation}, that was 
proposed by White \cite{re:white96} first
in the context of providing a guess for the initial Lanczos vector to speed up the algorithm, but
later found to be of applicability for other sub-algorithms.

After the state $|\phi\rangle$ in Eq.~(\ref{eq:b}) is computed, the DMRG algorithm operates for a few extra steps to better
converge all states $|\psi(t_x)\rangle$, $\forall t_x\le \tau$. These states
and all DMRG transformations can be saved to disk, and later be
used to compute the observables Eq.~(\ref{eq:generalcorrelation}).

\subsection{Evolution to Arbitrary Times}\label{sec:timeevolution}

What happens to Eq.~(\ref{eq:generalcorrelation}) for larger times, \ie, for times $t>\tau$?
Noting that $|\phi(\tau+\tau)\rangle=\exp(iH\tau)|\phi(\tau)\rangle$ we can apply Eq.~(\ref{eq:lanczosevolution})
to $|\phi(\tau)\rangle$, which in general reads~\cite{re:manmana05}:
\begin{equation}
|\phi(t+\tau)\rangle_i = \sum_{k,k',l,l',j} V^\dagger_{i,k} T^\dagger_{k,k'} S^\dagger_{k',l'}
{\rm e}^{i d_{l'} \tau} S_{l',l} T_{l,0} V_{0,j} |\phi(t)\rangle_j.
\label{eq:lanczosevolution2}
\end{equation}
Then, we proceed by \emph{targetting} the
states $\{|\phi(t_x+\tau)\rangle\}_x$ for some time until they are converged.
By applying this procedure recursively, we reach arbitrary times $t$ as we sweep the finite lattice
back and forth, and target $\{|\phi(t_x+t)\rangle\}_x$ in the general case.

The speed of time advancement in the algorithm is controlled by two opposing computational constraints.
If we advance times too fast by applying Eq.~(\ref{eq:lanczosevolution2}) too often, then
convergence might not be achieved, or we might not have had the chance to visit all sites
$\pi(0)$, $\pi(1)$, \ldots, $\pi(a-1)$ to compute Eq.~(\ref{eq:generalcorrelation}).
Conversely, advancing too slowly would increase computational cost but produce no
additional data.
Remember that when not advancing in time, states ${|\phi(t+t_x)\rangle}_x$
 are \emph{wave-function-transformed}, as explained in the previous section.

We now explain the choice of the weights~\cite{re:batrouni06} that appear in Eq.~(\ref{eq:rdensitymatrix}).
Assume $n$ to be the total number of states to be targetted, including the ground state.
To give them more prominence, we have chosen a weight of $2\Omega$ for the ground state,
and also for the $\Phi(l)$ vectors at
the beginning and end of the $\tau$ interval. We have chosen a weight of $\Omega$ for the rest.
Then  $2\Omega\times 3 + \Omega (n-3) = 1$ implies $\Omega = 1/(n+3)$.
The algorithm does not appear much dependent on the choice of weights.
However, irrespective of what the choice actually is, 
all mentioned states must have non-zero weights to
avoid loss of precision for one or more states.

\subsection{Overview of the Implementation}\label{sec:implementation}
The DMRG++ code was introduced in Refs.~\cite{re:alvarez09,re:alvarez11}.
The extension of the code to handle the time evolution and computation of observables of the type represented
by Eq.~(\ref{eq:generalcorrelation}) was carried out with minimal refactoring.
A \emph{Targetting} interface was introduced, with two concrete classes, \verb=GroundStateTargetting=,
and \verb=TimeStepTargetting=. The first handles the usual case, and is used even in the presence of time
evolution during the so-called ``infinite'' DMRG algorithm, and during the
finite algorithm before encountering the first site $\pi'(0)$ in Eq.~(\ref{eq:b}).

A call to \verb=target.evolve(...)= handles (i) the computation of the vectors $\{|\phi(t+t_x)\rangle\}_x$
as needed, and (ii) their time evolution or,
depending on the stage of the algorithm, their \emph{wave-function-transformation}.
When the \verb=target= object
belongs to the \verb=TimeStepTargetting= class, the actual implementation of these
tasks is performed by the member function \verb=evolve(...)=. When  the \verb=target= object
is of class \verb=GroundStateTargetting= the \verb=evolve(...)= function is empty.
The call to this function is always done immediately after obtaining the ground state $|\psi\rangle$
for that particular step of the DMRG algorithm.
	
File \verb=TimeStepTargetting.h= is documented in place using literate programming~\cite{re:knuth92}.
Further details about how to run the DMRG++ code, and how to 
specify its input file are given in
Appendix~\ref{app:computercode}.

\section{Examples}\label{sec:casestudies}

\subsection{One-site Excitations}\label{sec:onesiteexciton}

To test the accuracy of the time-dependent DMRG approach explained in the
previous section, we consider first the following problem. Consider 
the tight-binding model $H_0=\sum_{i,j,\sigma} t_{ij} c^\dagger_{i\sigma} c_{j\sigma}$, with
$t_{ij}$ a symmetric matrix, and with the observable we wish to calculate being
\begin{equation}
X_{ij\uparrow}(t)=\langle\psi| c^\dagger_{i\uparrow} {\rm e}^{-iHt} n_{j\uparrow}
{\rm e}^{iHt} c_{i\uparrow} |\psi\rangle,
\label{eq:xij}
\end{equation}
where $|\psi\rangle$ is the ground state of $H_0$.
(We keep the
usual notation $t_{ij}$ for the matrix of hopping integrals in the context of a
tight binding model~\cite{re:slater54}, even though $t$ is also used to denote
time here.)

This is equivalent to taking $b=1$,  $B=c_{\uparrow}$, and $\pi'(0)=i$ in Eq.~(\ref{eq:b});
and $a=1$, $\pi(0)=j$, and $A=n_{\uparrow}$ in Eq.~(\ref{eq:generalcorrelation}).
The physical interpretation for $X_{ij\uparrow}(t)$ is then clear:  it provides
the time-dependent expectation value of the charge density $\langle
n_{j\uparrow} \rangle (t)$ at site $j$ over a state that, at time $t=0$, is
defined by creating  a ``hole-like" excitation in state $|\psi\rangle$ at site
$i$.
 We assume that site $i$ has been specified and is kept fixed
throughtout this discussion.

\begin{figure}[t]
\includegraphics[width=8cm,clip]{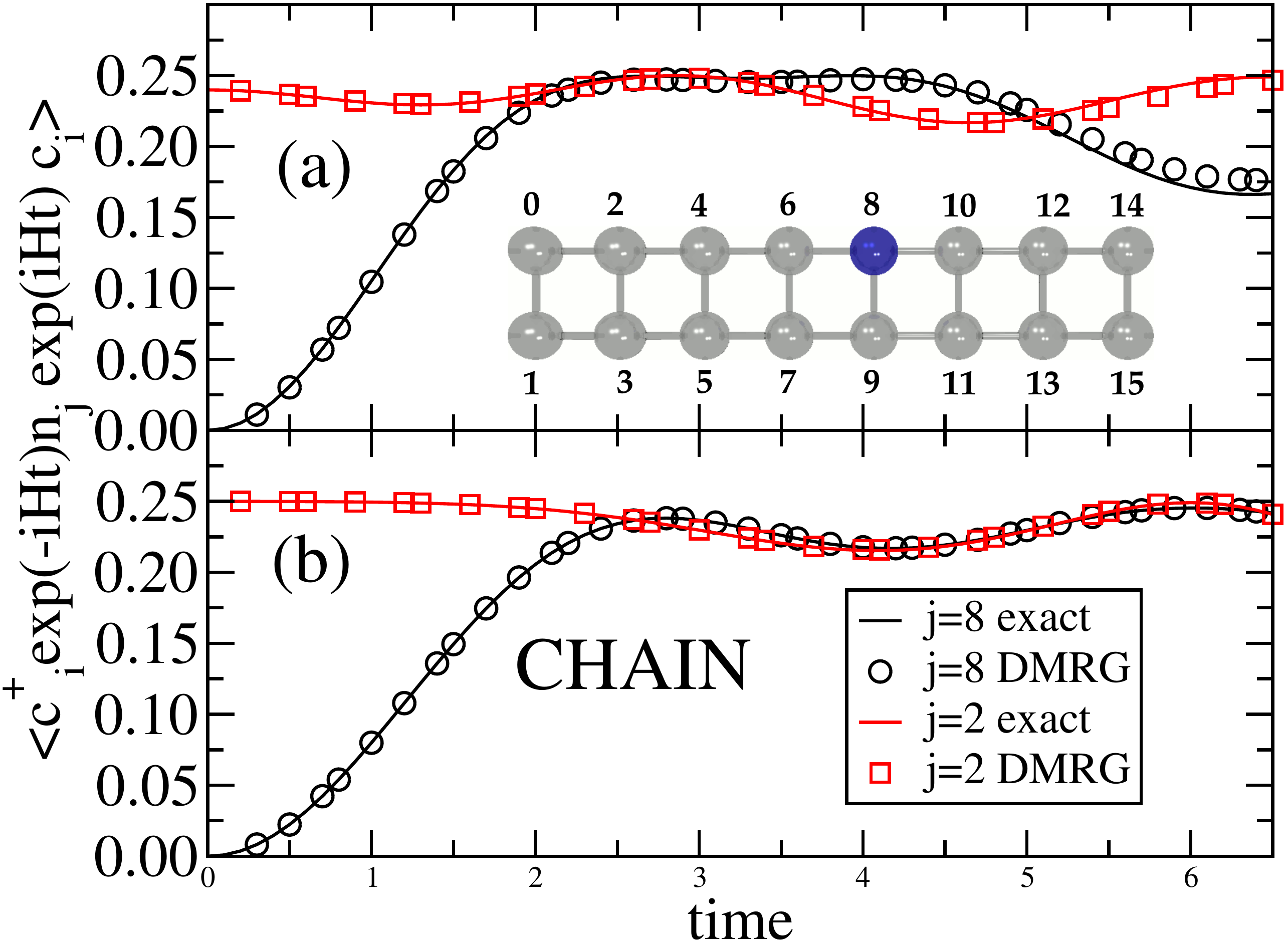}
\caption{\label{fig:output8cl} Observable $X_{ij\uparrow}(t)$, Eq.~(\ref{eq:xij}),
for Hamiltonian $H_0$, with $i=8$ and $j$ as indicated, on
two geometries: (a) a 8$\times$2 ladder, and (b) a 16-site chain.
Circles and squares represent DMRG results, and solid lines exact
results.
Inset: Labelling of sites on a two-leg ladder.
The highlighted site $i=8$ is where the one-site excitation (creation of
a ``hole'') was applied.
Chain sites are labelled from left to right, starting at 0.
\label{fig:easyexciton}}
\end{figure}

$X_{ij}(t)$ can be expressed in terms of the eigenvectors and eigenvalues of
$t_{ij}$. For a half-filled lattice we have $X_{ij}(t) = R_iR_j -
|T_{ij}(t)|^2$, where $R_i$ and $T_{ij}$ are given in
Appendix~\ref{app:easyexciton}. Then, $X_{ij}(t)$ can be calculated numerically,
and compared to DMRG results for this model on a 16-site chain, and on a
8$\times$2 ladder, see Fig.~\ref{fig:easyexciton}.

On the chain, the number of states kept for the DMRG algorithm was set to
$m=200$, which was found to give good accuracy for the ground state energy. On
the ladder, $m=400$ was used, which is a typical~\cite{re:landau94} $m$ value to
achieve good accuracy for the ground state energy, and static properties.
For instance, in both cases, DMRG gives $X_{ii}(t=0)=0$ and $X_{ij}(t=0)=1/4$
for $i \neq j$, as expected.
As shown in Fig.~\ref{fig:easyexciton}, the use of these values of $m$ yields
an accurate time evolution.

%

\subsection{Holon-Doublon}\label{sec:holondoublon}
Currently there is considerable interest in studying the feasibility of a new class of materials---the Mott insulators--- for their possible use in 
photovoltaic devices and oxide-electronics in general.
The crucial question under study is whether
charge excitations in the Mott insulator will be able to properly transfer the charge into the metallic contacts,
thus establishing a steady-state photocurrent.
Answering this question will require computation of the out-of-equilibrium dynamics and the time evolution
of the excitonic excitations produced by the absorption of light by the
material.

The electron and hole created by light absorption 
are modeled by the state~\cite{re:al-hassanieh08}
\begin{equation}
|\phi\rangle\equiv c_{i\sigma}(1-n_{i\bar{\sigma}}) c^\dagger_{j\sigma'}n_{j\bar{\sigma'}} |\psi\rangle,
\label{eq:holondoublon}
\end{equation}
where $|\psi\rangle$ is the ground state,
$\sigma$ and $\sigma'$ are spin indices, and
$\bar{\sigma}=1-\sigma$ denotes the
 spin opposite to $\sigma$.
A sum over $\sigma$ and $\sigma'$ is assumed in the equation above.
This is equivalent to taking $b=2$,
$B_0=c^\dagger_{\sigma}n_{\bar{\sigma'}}$,
$B1=c_{\sigma}(1-n_{\bar{\sigma}})$,
$\pi'(0)=j$, and $\pi'(1)=i$
in Eq.~(\ref{eq:b}).
We assume that the sites $i$ and $j$ of the lattice have been specified and will remain fixed
throughout this discussion.

To model a Mott insulator
we consider the Hubbard Hamiltonian~\cite{re:hubbard63,re:hubbard64a,re:hubbard64b}
\begin{equation}
\hat{H}=\hat{H}_0
+U\sum_i \hat{n}_{i\uparrow}\hat{n}_{i\downarrow},
\end{equation}
where the notation is as in Ref.~\cite{re:al-hassanieh08}, and we will drop
the hat from the operators from now on.
The hopping matrix $t$ corresponds either to an open chain 
or to a two-leg ladder in the studies below.

\subsubsection{Density}\label{sec:hd-dens}
The time-dependent density at site $p$ of state Eq.~(\ref{eq:holondoublon}) is
\begin{equation}
O_{j,i,p,\uparrow}(t) = \frac{\langle \phi|{\rm e}^{-iHt} n_{p\uparrow} {\rm e}^{iHt} | \phi\rangle
}{\langle\phi |\phi\rangle},
\label{eq:nholondoublon}
\end{equation}
which amounts to taking  $a=1$,
$A=n_{\uparrow}$, and $\pi(0)=p$ in  Eq.~(\ref{eq:generalcorrelation}).
Consider $O_{j,i,p}(t) = \sum_\sigma O_{j,i,p,\sigma}(t)$.
In the case of $U=0$ and half-filling we have (details are in Appendix~\ref{app:holondoublon}):
\begin{equation}
O_{j,i,p}(0)\equiv O_{j,i,p,\uparrow}(0) + O_{j,i,p,\downarrow}(0)=
\left\{
\begin{tabular}{ll}
$0$ & if $p=i$\\
$2$ & if $p=j$\\
$1$ & otherwise.
\end{tabular}
\right.
\label{eq:nd}
\end{equation}

\begin{figure}
\includegraphics[width=8cm,clip]{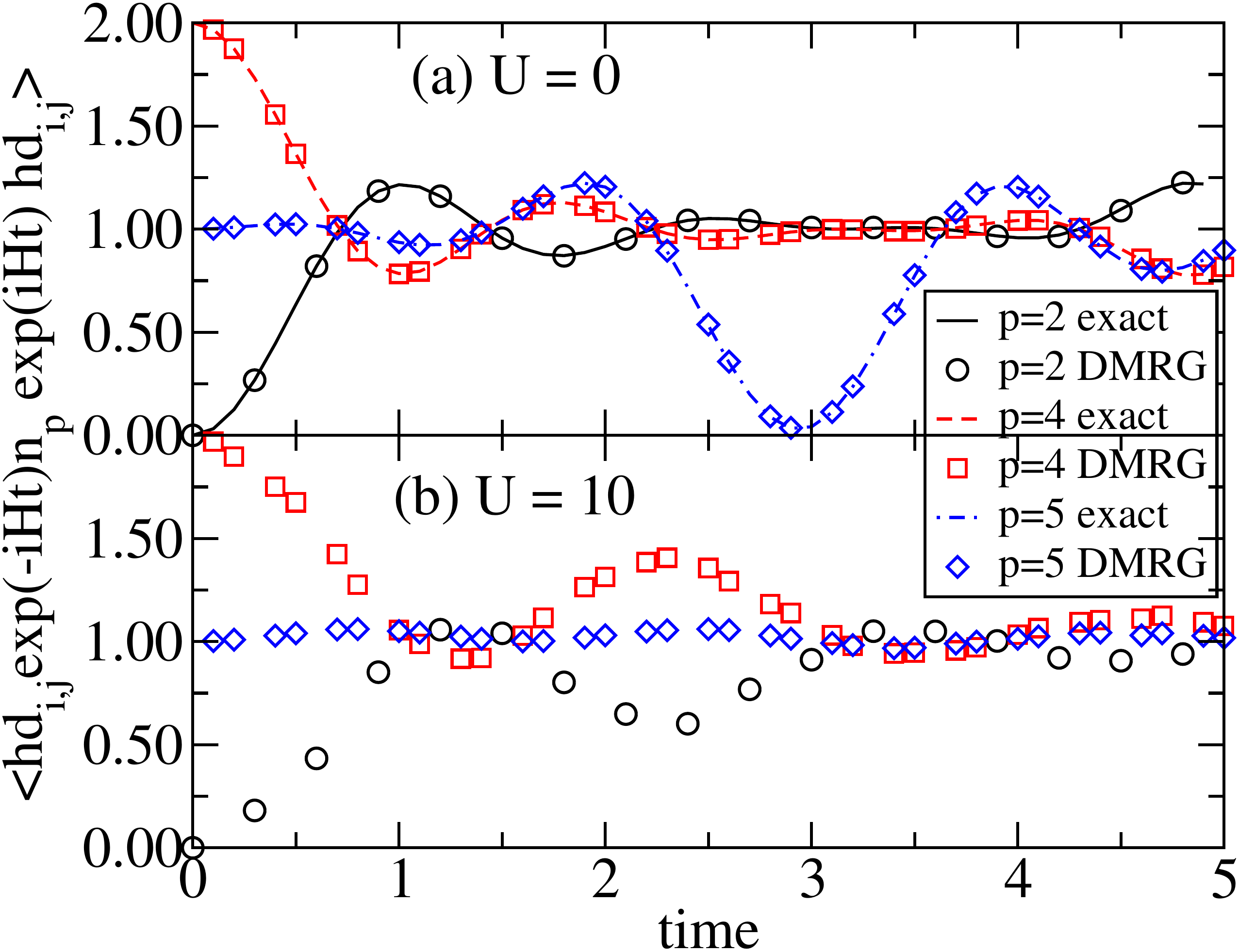}
\caption{\label{fig:holonDoublonDensity} Density at site $p$ of the
holon-doublon state Eq.~(\ref{eq:holondoublon}) as a function of time, for (a) $U=0$, and (b) $U=10$.
A 2$\times$4 ladder with $t_x=1$ and $t_y=0.5$ was used, containing 4 up and 4 down electrons.
The holon-doublon operator was applied at $i=2$, $j=4$.}
\end{figure}

The observable we test in this section is $\langle \Psi_e|n_p|\Psi_e\rangle$,
which has a similar physical interpretation as  $X_{ij}(t)$
(Eq.~(\ref{eq:xij})) in the holon-doublon case.
Results for $U=0$ and $U=10$ are shown in Fig.~\ref{fig:holonDoublonDensity}.
At $t=0$, the values of $\langle \Psi_e|n_p|\Psi_e\rangle$ given by
Eq.~(\ref{eq:nd}) hold true in the $U\neq0$ case at half-filling. The
time-evolution for interacting and non-interacting cases are, however, quite
distinct, as in the case of the chain (see, e.g.,
Ref.~\cite{re:diasdasilva10}).

Readers might want to know why we emphasize the non-interacting $U=0$ case. One
obvious advantage of the $U=0$ case is that we can test the Krylov method, and
indirectly the accuracy of the DMRG, against exact results. In addition, the $U$ term, at least
when on-site, is not a major source of efficiency problems for the DMRG
algorithm.

\begin{figure}
\centering{
\includegraphics[width=8cm,clip]{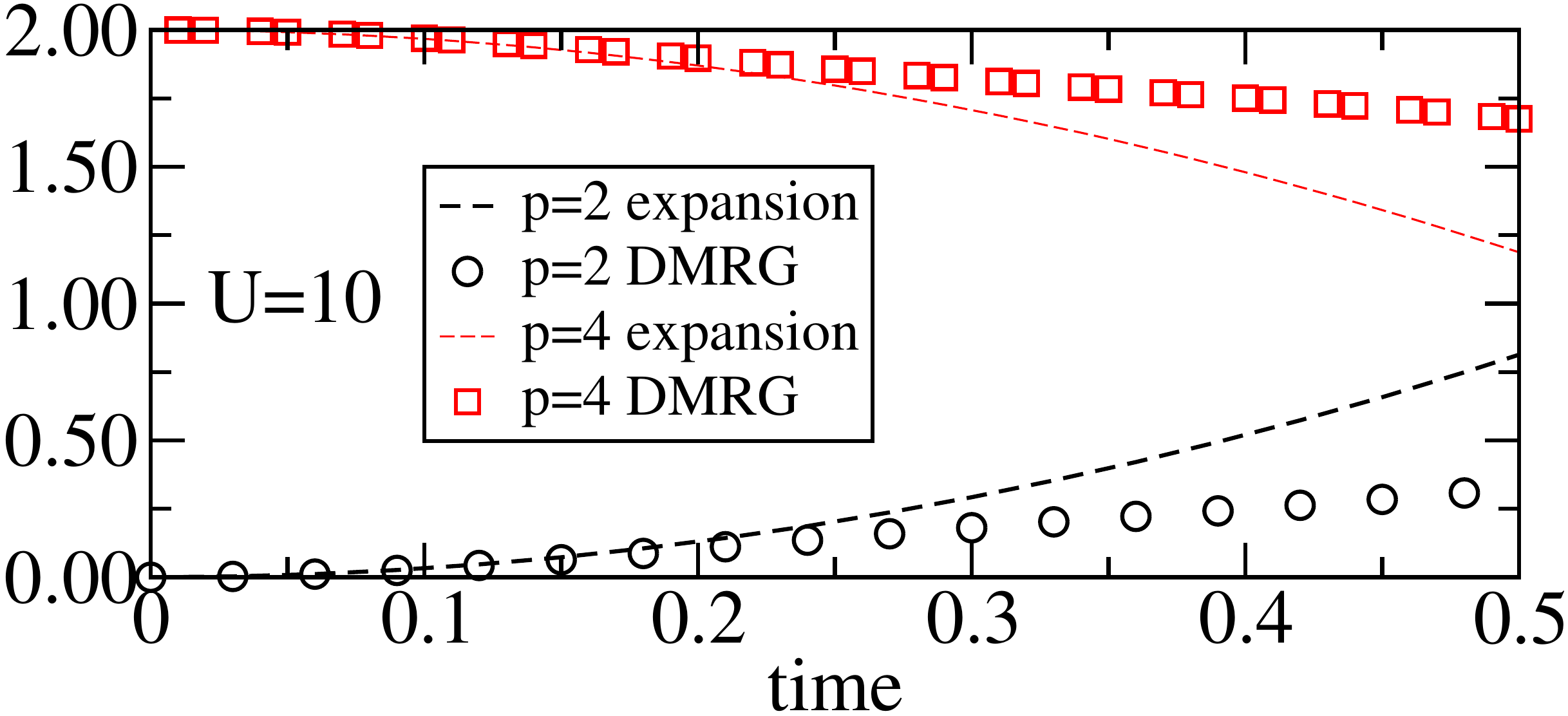}
}
\caption{\label{fig:smalltime} Comparison of $O_{j,i,p}(t)$ computed with DMRG (circles and squares)
and with a $t\rightarrow0$ expansion. For the latter $O_{j,i,p}(t)=O_{j,i,p}(0)+a_{j,i,p}t^2+O(t^4)$,
where $a_{j,i,p}\equiv\langle\phi|Hn_p H - H^2n_p|\phi\rangle$.
Same parameters as in Fig.~\ref{fig:holonDoublonDensity}b.
}
\end{figure}

To test our results for $U\neq0$ we have compared them to the Suzuki-Trotter method (not shown).
We have also computed the small time expansion, and this is shown in Fig.~\ref{fig:smalltime}.

\subsubsection{Double-occupation}\label{sec:nd}

The double-occupation of state Eq.~(\ref{eq:holondoublon}) is~\cite{re:al-hassanieh08}
\begin{equation}
N_d(j,i,p,t) \equiv \frac{\langle \phi|
 {\rm e}^{-iHt}
n_{p\uparrow} n_{p\downarrow}
{\rm e}^{iHt}
|\phi \rangle}{\langle\phi|\phi\rangle},
\end{equation}
and amounts to taking  $a=1$,
$A=n_\uparrow n_\downarrow$, and $\pi(0)=p$ in  Eq.~(\ref{eq:generalcorrelation}).

%
Summarizing 
the \emph{operator} equations obtained in section~\ref{sec:hd-dens},
$n_{i\uparrow}A_{ij}=0$, and
$\bar{n}_{j\uparrow}A_{ij}=0$,
where $A_{ij}$ is the operator defined in Eq.~(\ref{eq:a}), $\bar{n}=1-n$, and these equations also
hold if we replace $\uparrow$ by $\downarrow$.
Then $N_d(j,i,i,t=0)\langle\phi|\phi\rangle= \langle A_{ij}^\dagger n_{i\uparrow}n_{i\downarrow}A_{ij}\rangle = 0$, and

\begin{multline}
N_d(j,i,j,t=0)\langle\phi|\phi\rangle = \langle A_{ij}^\dagger n_{j\uparrow} n_{j\downarrow} A_{ij}\rangle=\\
\langle A_{ij}^\dagger n_{j\uparrow} (1-\bar{n}_{j\downarrow})A_{ij}\rangle
=\langle A_{ij}^\dagger n_{j\uparrow} A_{ij}\rangle \equiv O_{j,i,j,\uparrow}\langle \phi|\phi\rangle\nonumber.
\end{multline}

\begin{figure}
\includegraphics[width=8cm,clip]{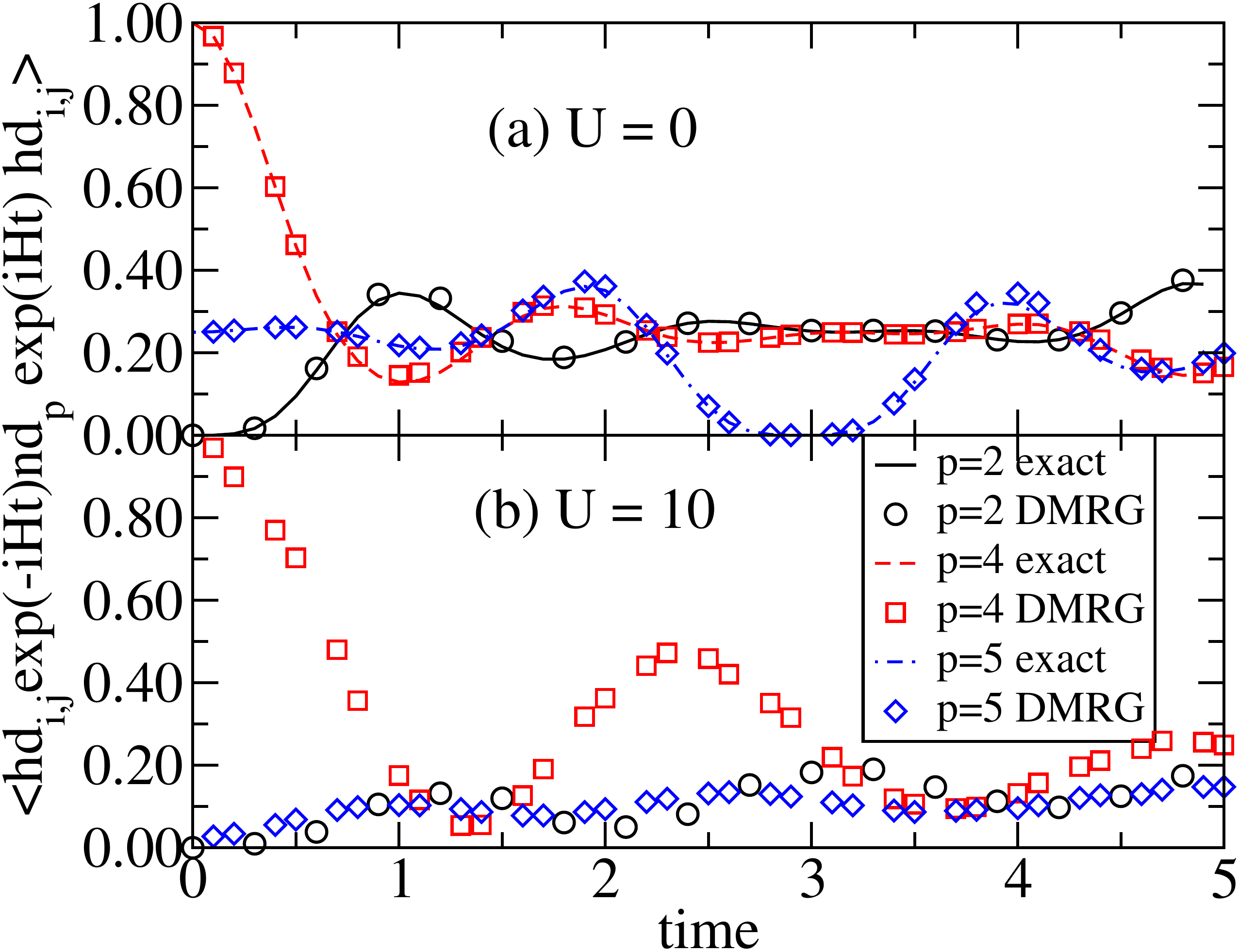}
\caption{\label{fig:holonDoublonNd} Double occupation at site $p$ of the
holon-doublon state Eq.~(\ref{eq:holondoublon}) as a function of time, for (a) $U=0$ and (b) $U=10$.
A 2$\times$4 ladder with $t_x=1$ and $t_y=0.5$ was used, containing 4 up and 4 down electrons.
The holon-doublon operator was applied at $i=2$, $j=4$.}
\end{figure}

DMRG results
for $U=0$ and $U=10$ are shown in Fig.~\ref{fig:holonDoublonNd}.
Also shown are exact results for $U=0$. At $t=0$, the double occupation at the
doublon ($p=j$) and holon ($p=i$) sites are, respectively, $N_d(j,i,p=j,t=0)=1$
and $N_d(j,i,p=i,t=0)=0$ for both interacting and noninteracting cases. At the
doublon site, the double occupation has a characteristic oscillating decay
caused by the dynamics of the holon-doublon pair within the system, also observed
for the chain case \cite{re:diasdasilva10}.

\section{Computational Efficiency and Concurrency}\label{sec:performance}
As in the static DMRG algorithm,
the most computationally intensive task of the time-step targetting DMRG algorithm is the
computation of Hamiltonian connections between the system and environment blocks.
The difference is that now the lattice needs to be swept painstakenly to advance to larger and
larger times. The scaling, however, is linear with the number of finite sweeps,
as long as the truncation $m$ remains constant.

This expensive task of building Hamiltonian connections between system and environment blocks can be
parallelized~\cite{re:hager04,re:chan04,re:kurashige09,re:yamada09,re:rincon10}.
Our implementation uses \emph{pthreads}, a shared memory approach~\footnote{\emph{Pthreads} or
POSIX threads is a standardized C language threads programming interface,
 specified by the IEEE POSIX standard.}.
In percentage, the computation speed-up is similar  to the static DMRG case, and a discussion of the strong
scaling can be found in Ref.~\cite{re:alvarez11}. In terms of wall-clock time, the speed-up is larger due to the
time-step targetting DMRG algorithm taking more time than the static version.

The computation of target states could be parallelized easily, but whether serial or parallel,
it is too fast to have substantial impact on the CPU times of production runs.

The measurement of observables is a different matter. DMRG++ computes observables \emph{post-processing}, \ie,
the main code saves all DMRG transformations, permutations, and quantum states to disk, and a
second \emph{observer} code reads the data from disk and computes observables as needed.
We argue that \emph{post-processing} is more advantageous than \emph{in-situ processing}, whether for the static
or for the time-dependent DMRG algorithm.

First, a single run of the main code enables computation of all observables.
If, instead, one needed to make a decision on what observables to compute when
running the main code, one would risk computing too much or too little. In the former case, computational
resources and wall clock time would be wasted. In the latter case, the main run would have to be repeated, leading to vast
redunduncies because the observations are not computationally intensive compared to the main code.

Moreover, by computing observables \emph{post-processing}, we decouple the code, and enable scalable parallel computations.
For example, one-point observables of the form  $\langle \phi_1(t) | A_i |\phi_2(t)\rangle$
are parallelized over $i$, and  two-point
correlations, such as $\langle \phi_1(t) | A_i B_j |\phi_2(t)\rangle$, are parallelized over $j$,
and cached over $i$.
If $N$ is the
number of sites of the lattice, the parallelization scales linearly up to almost $N$; the scaling is good but not perfect due to
initialization costs~\cite{re:amdahl67}.

\section{Summary}\label{sec:summary}
This paper explained in detail the implementation of the
Krylov method for the real time evolution within the DMRG algorithm, using time-step
targetting~\cite{re:batrouni06,re:manmana05}. We applied the method to a simple case of one-site excitations and found
the method to be accurate. For the case of the holon-doublon excitation,
we have extended to two-leg ladders the 
previous results obtained in chains. Our analysis has shown 
that the method is accurate as long as
the underlying DMRG algorithm is accurate. Since Mott insulators are under study for its possible
applicability to solar cells, the present results pave the way for their continued study, now on more complex (but still quasi-one dimensional) geometries, such as ladders.

We described computational tricks that can help decrease the 
runtime. For example, we mentioned that
shared memory parallelization with a few CPU cores can cut times by a factor of 2 or more.
Parallelization works in the same way for time-dependent DMRG as it does for static DMRG, but
helps more in the former case, due to runs taking longer.
We also argued in favor of the \emph{post-processing} of observables to speed-up production runs,
and increase computational efficiency.

Our implementation, DMRG++, is free and open source.
It emphasizes generic programming using C++ templates, a 
friendly user-interface, and as few software dependencies as possible.
DMRG++ makes writing new models and geometries easy and fast, by using a generic DMRG engine.

\appendix
\section{One-site Excitation in the Non-Interacting Case}\label{app:easyexciton}
Let $U$ be the matrix that diagonalizes $H_0$, so that
$c^\dagger_{i\uparrow}  = \sum_\lambda U^*_{i,\lambda,\uparrow} u_{\lambda,\uparrow}$,
and $u_{\lambda,\uparrow}$ the diagonal operators. Let $E_\lambda$ be the eigenvalues of $H_0$.

For the rest of this appendix we omit $\uparrow$.
After some algebra, and omitting the sums over duplicated indices:
\begin{equation}
X_{ij}(t)=U^*_{i,\lambda}U^*_{j,\xi}U_{j,\xi'}U_{i,\lambda'}\langle
u^\dagger_\lambda {\rm e}^{-iHt} u^\dagger_\xi u_\xi {\rm e}^{iHt} u_{\lambda'}\rangle.
\end{equation}
The ground state of $H_0$ is made up of $N_\uparrow$ filled levels, up to
the Fermi energy, and particle-hole excitations are
eigenstates of $H$ (or, conversely, the excited states of $H_0$ are particle-hole excitations).
Then, the $\lambda'-$th level of $|\phi\rangle$ is occupied
and $u_{\lambda'}|\phi\rangle$ is an eigenstate of $H$ with a hole at $\lambda'$.
Applying this reasoning multiple times,
%
the final result is $X_{ij}(t) = R_iR_j - |T_{ij}(t)|^2$, where
$R_i = \sum^{'}_{\lambda} |U_{i,\lambda}|^2$, and
$T_{ij}(t) = \sum^{'}_{\lambda}U^{*}_{i,\lambda} U_{j,\lambda} \exp(-iE_\lambda t)$,
where the prime over the sumation means sum only over occupied states $\lambda$.


\section{Holon-Doublon for a Non-Interacting System}\label{app:holondoublon}
 The goal of this appendix is to compute Eq.~(\ref{eq:nholondoublon}) when $U=0$ and $i\neq j$.
Let $N$ be the number of sites,
let $N$ be even, and let there be $N_{\uparrow}=N_{\downarrow}=N/2$ electrons.
Then $\langle n_{i\uparrow}\rangle = \langle n_{i\downarrow}\rangle=\frac 12$,
$\forall i$.

Let $\mathcal{N}_{ij\uparrow}\equiv\langle n_{i\uparrow} n_{j\uparrow}\rangle$
be an equation between real numbers, and $n_i \equiv n_{i\uparrow}+n_{i\downarrow}$ be an operator equation.
From $\langle n_{i\uparrow}\rangle = \langle n_{i\downarrow}\rangle=\frac12$, it follows that
$2\mathcal{N}_{ij\uparrow}+\frac 12 =\langle n_i n_j\rangle$.
Another straightforward result is $O_{j,i,i,\uparrow}(0)=0$.

For an operator $A$ let us define $\mathcal{D}(A)=\langle A^\dagger A \rangle$, and
let
\begin{equation}
A_{ij}= c_{i\sigma}(1-n_{i\bar{\sigma}})c^\dagger_{j\sigma'}n_{j\bar{\sigma'}},
\label{eq:a}
\end{equation}
then after some algebra:
\begin{equation}
\mathcal{D}(A_{ij})\equiv \mathcal{D}_{ij}=\frac 12 -
2\mathcal{N}_{ij\uparrow}+4\mathcal{N}_{ij\uparrow}^2 + 2|\mathcal{X}_{ij\uparrow}|^2,
\label{eq:dij}
\end{equation}
where $\mathcal{X}_{ij\uparrow}=\langle c^\dagger_{i\uparrow}c_{j\uparrow}\rangle$.

By writing $n_{p\uparrow} = 1-c_{p\uparrow} c^\dagger_{p\uparrow}$, one gets $O_{i,j,j,\uparrow}(0)=1- {\rm term}$.
It is straightforward to prove that the second term vanishes and thus:
$O_{j,i,j,\uparrow}(0)=1$.

Now consider $p\neq i$ and $p\neq j$. If we expand in the basis that diagonalizes the Hamiltonian we arrive to:
\begin{equation}
O_{j,i,p,\uparrow}(0)=\langle n_{p\uparrow}\rangle
+\sum_{n>0} \langle A^\dagger_{ij} A_{ij}|n\rangle\langle n| n_{p\uparrow}\rangle/\mathcal{D}_{ij},
\label{eq:ojip}
\end{equation}
where the sum in $n$ is over excited states. The second term is non-negative.
This follows because $A^\dagger_{ij} A_{ij}$ and $n_{p\uparrow}$ commute, admiting a common basis where both are diagonal.
Moreover, all eigenvalues of $A^\dagger_{ij} A_{ij}$ are non-negative, as are all those of $n_{p\uparrow}$.
Then, the second term in Eq.~(\ref{eq:ojip}) is non-negative.

Also, the first term of Eq.~(\ref{eq:ojip}) is $\frac 12$.  Then $\forall p\neq i,j$
$O_{j,i,p,\uparrow}(0)=\frac12 + r_{ijp}$,
where $r_{ijp}\ge 0$.

If we sum over all sites,
\begin{equation}
\langle A^\dagger_{ij} A_{ij} \sum_r n_{r\uparrow}\rangle/\mathcal{D}_{ij} =
\langle A^\dagger_{ij} A_{ij} N_{\uparrow}\rangle/\mathcal{D}_{ij}=\frac N2.
\label{eq:oijpsum}
\end{equation}
There are $N-2$ sites  $p$  such that $p\neq i,j$. Putting it all together we get:
 \begin{equation}
\sum_r O_{j,i,r,\uparrow}(0) = 1 + \frac{N-2}{2} +\sum_{p\neq i,j} r_{ijp}=
\frac{N}{2} +\sum_{p\neq i,j} r_{ijp},
\end{equation}
which, according to Eq.~(\ref{eq:oijpsum}) has to be equal to $\frac N2$.
It follows that $\sum_{p\neq i,j} r_{ijp}=0$,
implying $r_{ijp}=0$ $\forall p\neq i,j$ since we knew the $r$s were non-negative.
Hence $O_{j,i,p,\uparrow}(0)=\frac12$, $\forall p\neq i,j$

\section{The DMRG++ Code}\label{app:computercode}
The required software to build DMRG++ is:
(i) GNU C++, and
(ii) the LAPACK and BLAS libraries~\cite{laug}.
These libraries are available for most platforms.
The configure.pl script will ask for the \verb=LDFLAGS= variable
to pass to the compiler/linker. If the \emph{Linux} platform was
chosen the default/suggested \verb=LDFLAGS= will include \verb=-llapack=.
If the \emph{OSX} platform was chosen the default/suggested \verb=LDFLAGS= will
include  \verb=-framework Accelerate=.
For other platforms the appropriate linker flags must be given.
More information on \verb=LAPACK= is here: http://netlib.org/lapack/.

Optionally, make or gmake is needed to use the Makefile, and perl
is only needed to run the \verb=configure.pl= script.

To build and run DMRG++:
\begin{verbatim}
cd src
perl configure.pl
(please answer questions regarding model, and
choose TimeStepTargetting)
make dmrg; make observe
./dmrg input.inp
./observe input.inp time
\end{verbatim}

The perl script \verb=configure.pl= will create the
files
\verb=main.cpp=, \verb=Makefile= and
\verb=observe.cpp=.
Example input files for one-site excitations are in
\verb=TestSuite/inputs/input8.inp=, and for holon-doublon excitations in
\verb=TestSuite/inputs/input10.inp=.
These files can be modified and used as input to run the DMRG++ program.
Further details can be found in the file README in the code.

The \verb=FreeFermions= code found at \url{http://www.ornl.gov/~gz1/FreeFermions/} can be
used to compute properties of non-interactive systems.

\begin{acknowledgments}
This work was supported by the
Center for Nanophase Materials Sciences, sponsored by the Scientific User Facilities Division, Basic Energy Sciences, U.S.
Department of Energy,
under contract with UT-Battelle.
This research used resources of the National Center for Computational Sciences, as well as
the OIC at Oak Ridge National Laboratory. E.D. is supported by the U.S. Department of Energy, Office of Basic Energy
Sciences, Materials Sciences and Engineering Division.\end{acknowledgments}

\bibliography{thesis}

\begin{thebibliography}{42}
\expandafter\ifx\csname natexlab\endcsname\relax\def\natexlab#1{#1}\fi
\expandafter\ifx\csname bibnamefont\endcsname\relax
  \def\bibnamefont#1{#1}\fi
\expandafter\ifx\csname bibfnamefont\endcsname\relax
  \def\bibfnamefont#1{#1}\fi
\expandafter\ifx\csname citenamefont\endcsname\relax
  \def\citenamefont#1{#1}\fi
\expandafter\ifx\csname url\endcsname\relax
  \def\url#1{\texttt{#1}}\fi
\expandafter\ifx\csname urlprefix\endcsname\relax\def\urlprefix{URL }\fi
\providecommand{\bibinfo}[2]{#2}
\providecommand{\eprint}[2][]{\url{#2}}

\bibitem[{\citenamefont{Heidrich-Meisner
  et~al.}(2010)\citenamefont{Heidrich-Meisner, Gonzalez, Al-Hassanieh, Feiguin,
  Rozenberg, and Dagotto}}]{re:heidrich-meisner10}
\bibinfo{author}{\bibfnamefont{F.}~\bibnamefont{Heidrich-Meisner}},
  \bibinfo{author}{\bibfnamefont{I.}~\bibnamefont{Gonzalez}},
  \bibinfo{author}{\bibfnamefont{K.}~\bibnamefont{Al-Hassanieh}},
  \bibinfo{author}{\bibfnamefont{A.}~\bibnamefont{Feiguin}},
  \bibinfo{author}{\bibfnamefont{M.}~\bibnamefont{Rozenberg}},
  \bibnamefont{and} \bibinfo{author}{\bibfnamefont{E.}~\bibnamefont{Dagotto}},
  \bibinfo{journal}{Phys. Rev. B} \textbf{\bibinfo{volume}{104508}},
  \bibinfo{pages}{82} (\bibinfo{year}{2010}).

\bibitem[{\citenamefont{da~Silva et~al.}(2009)\citenamefont{da~Silva, Tiago,
  Ulloa, Reboredo, and Dagotto}}]{re:diasdasilva09}
\bibinfo{author}{\bibfnamefont{L.~G. G. V.~D.} \bibnamefont{da~Silva}},
  \bibinfo{author}{\bibfnamefont{M.~L.} \bibnamefont{Tiago}},
  \bibinfo{author}{\bibfnamefont{S.~E.} \bibnamefont{Ulloa}},
  \bibinfo{author}{\bibfnamefont{F.~A.} \bibnamefont{Reboredo}},
  \bibnamefont{and} \bibinfo{author}{\bibfnamefont{E.}~\bibnamefont{Dagotto}},
  \bibinfo{journal}{Phys. Rev. B} \textbf{\bibinfo{volume}{80}},
  \bibinfo{pages}{155443} (\bibinfo{year}{2009}).

\bibitem[{\citenamefont{Heidrich-Meisner
  et~al.}(2009{\natexlab{a}})\citenamefont{Heidrich-Meisner, Manmana, Rigol,
  Muramatsu, Feiguin, and Dagotto}}]{re:heidrich-meisner09}
\bibinfo{author}{\bibfnamefont{F.}~\bibnamefont{Heidrich-Meisner}},
  \bibinfo{author}{\bibfnamefont{S.~R.} \bibnamefont{Manmana}},
  \bibinfo{author}{\bibfnamefont{M.}~\bibnamefont{Rigol}},
  \bibinfo{author}{\bibfnamefont{A.}~\bibnamefont{Muramatsu}},
  \bibinfo{author}{\bibfnamefont{A.~E.} \bibnamefont{Feiguin}},
  \bibnamefont{and} \bibinfo{author}{\bibfnamefont{E.}~\bibnamefont{Dagotto}},
  \bibinfo{journal}{Phys. Rev. A} \textbf{\bibinfo{volume}{041603(R)}},
  \bibinfo{pages}{80} (\bibinfo{year}{2009}{\natexlab{a}}).

\bibitem[{\citenamefont{Heidrich-Meisner
  et~al.}(2009{\natexlab{b}})\citenamefont{Heidrich-Meisner, Feiguin, and
  Dagotto}}]{re:heidrich-meisner09b}
\bibinfo{author}{\bibfnamefont{F.}~\bibnamefont{Heidrich-Meisner}},
  \bibinfo{author}{\bibfnamefont{A.~E.} \bibnamefont{Feiguin}},
  \bibnamefont{and} \bibinfo{author}{\bibfnamefont{E.}~\bibnamefont{Dagotto}},
  \bibinfo{journal}{Phys. Rev. B} \textbf{\bibinfo{volume}{235336}},
  \bibinfo{pages}{79} (\bibinfo{year}{2009}{\natexlab{b}}).

\bibitem[{\citenamefont{Heidrich-Meisner
  et~al.}(2009{\natexlab{c}})\citenamefont{Heidrich-Meisner, Martins,
  B{\"u}sser, Al-Hassanieh, Feiguin, Chiappe, Anda, and
  Dagotto}}]{re:heidrich-meisner09c}
\bibinfo{author}{\bibfnamefont{F.}~\bibnamefont{Heidrich-Meisner}},
  \bibinfo{author}{\bibfnamefont{G.}~\bibnamefont{Martins}},
  \bibinfo{author}{\bibfnamefont{C.}~\bibnamefont{B{\"u}sser}},
  \bibinfo{author}{\bibfnamefont{K.}~\bibnamefont{Al-Hassanieh}},
  \bibinfo{author}{\bibfnamefont{A.}~\bibnamefont{Feiguin}},
  \bibinfo{author}{\bibfnamefont{G.}~\bibnamefont{Chiappe}},
  \bibinfo{author}{\bibfnamefont{E.}~\bibnamefont{Anda}}, \bibnamefont{and}
  \bibinfo{author}{\bibfnamefont{E.}~\bibnamefont{Dagotto}},
  \bibinfo{journal}{Eur. Phys. J.} \textbf{\bibinfo{volume}{527}},
  \bibinfo{pages}{67} (\bibinfo{year}{2009}{\natexlab{c}}).

\bibitem[{\citenamefont{White}(1992)}]{re:white92}
\bibinfo{author}{\bibfnamefont{S.}~\bibnamefont{White}},
  \bibinfo{journal}{Phys. Rev. Lett.} \textbf{\bibinfo{volume}{69}},
  \bibinfo{pages}{2863} (\bibinfo{year}{1992}).

\bibitem[{\citenamefont{White}(1993)}]{re:white93}
\bibinfo{author}{\bibfnamefont{S.}~\bibnamefont{White}},
  \bibinfo{journal}{Phys. Rev. B} \textbf{\bibinfo{volume}{48}},
  \bibinfo{pages}{345} (\bibinfo{year}{1993}).

\bibitem[{\citenamefont{Scholl{w\"o}ck}(2010)}]{re:schollwock10}
\bibinfo{author}{\bibfnamefont{U.}~\bibnamefont{Scholl{w\"o}ck}},
  \bibinfo{journal}{Annals of Physics} \textbf{\bibinfo{volume}{96}},
  \bibinfo{pages}{326} (\bibinfo{year}{2010}).

\bibitem[{\citenamefont{Schollw{\"o}ck}(2005{\natexlab{a}})}]{re:schollwock05}
\bibinfo{author}{\bibfnamefont{U.}~\bibnamefont{Schollw{\"o}ck}},
  \bibinfo{journal}{Rev. Mod. Phys.} \textbf{\bibinfo{volume}{77}},
  \bibinfo{pages}{259} (\bibinfo{year}{2005}{\natexlab{a}}).

\bibitem[{\citenamefont{Hallberg}(2006)}]{re:hallberg06}
\bibinfo{author}{\bibfnamefont{K.}~\bibnamefont{Hallberg}},
  \bibinfo{journal}{Adv. Phys.} \textbf{\bibinfo{volume}{55}},
  \bibinfo{pages}{477} (\bibinfo{year}{2006}).

\bibitem[{\citenamefont{Rodriguez-Laguna}(2002)}]{re:rodriguez02}
\bibinfo{author}{\bibfnamefont{J.}~\bibnamefont{Rodriguez-Laguna}}
  (\bibinfo{year}{2002}), \bibinfo{note}{http://arxiv.org/abs/cond-mat/0207340,
  Real Space Renormalization Group Techniques and Applications}.

\bibitem[{\citenamefont{Feiguin and White}(2005)}]{re:feiguin05}
\bibinfo{author}{\bibfnamefont{A.~R.} \bibnamefont{Feiguin}} \bibnamefont{and}
  \bibinfo{author}{\bibfnamefont{S.~R.} \bibnamefont{White}},
  \bibinfo{journal}{Phys. Rev. B} \textbf{\bibinfo{volume}{72}},
  \bibinfo{pages}{020404} (\bibinfo{year}{2005}).

\bibitem[{\citenamefont{Manmana et~al.}(2005)\citenamefont{Manmana, Muramatsu,
  and Noack}}]{re:manmana05}
\bibinfo{author}{\bibfnamefont{S.~R.} \bibnamefont{Manmana}},
  \bibinfo{author}{\bibfnamefont{A.}~\bibnamefont{Muramatsu}},
  \bibnamefont{and} \bibinfo{author}{\bibfnamefont{R.~M.} \bibnamefont{Noack}},
  in \emph{\bibinfo{booktitle}{AIP Conf. Proc.}} (\bibinfo{year}{2005}), vol.
  \bibinfo{volume}{789}, pp. \bibinfo{pages}{269--278}, \bibinfo{note}{also in
  http://arxiv.org/abs/cond-mat/0502396v1}.

\bibitem[{\citenamefont{Schollw{\"o}eck and White}(2006)}]{re:batrouni06}
\bibinfo{author}{\bibnamefont{Schollw{\"o}eck}} \bibnamefont{and}
  \bibinfo{author}{\bibnamefont{White}}, in \emph{\bibinfo{booktitle}{Effective
  models for low-dimensional strongly correlated systems}}, edited by
  \bibinfo{editor}{\bibfnamefont{G.~G.} \bibnamefont{Batrouni}}
  \bibnamefont{and} \bibinfo{editor}{\bibfnamefont{D.}~\bibnamefont{Poilblanc}}
  (\bibinfo{publisher}{AIP}, \bibinfo{address}{Melville, New York},
  \bibinfo{year}{2006}), p. \bibinfo{pages}{155}, \bibinfo{note}{also in
  http://de.arxiv.org/abs/cond-mat/0606018v1}.

\bibitem[{\citenamefont{Schollw{\"o}ck}(2005{\natexlab{b}})}]{re:schollwock05b}
\bibinfo{author}{\bibfnamefont{U.}~\bibnamefont{Schollw{\"o}ck}},
  \bibinfo{journal}{J. Phys. Soc. Jpn.} \textbf{\bibinfo{volume}{74}},
  \bibinfo{pages}{246} (\bibinfo{year}{2005}{\natexlab{b}}).

\bibitem[{\citenamefont{Daley et~al.}(2004)\citenamefont{Daley, Kollath,
  Schollw{\"o}ck, and Vidal}}]{re:daley04}
\bibinfo{author}{\bibfnamefont{A.~J.} \bibnamefont{Daley}},
  \bibinfo{author}{\bibfnamefont{C.}~\bibnamefont{Kollath}},
  \bibinfo{author}{\bibfnamefont{U.}~\bibnamefont{Schollw{\"o}ck}},
  \bibnamefont{and} \bibinfo{author}{\bibfnamefont{G.}~\bibnamefont{Vidal}},
  \bibinfo{journal}{J. Stat. Mech.: Theory Exp.} p. \bibinfo{pages}{P04005}
  (\bibinfo{year}{2004}).

\bibitem[{\citenamefont{White and Feiguin}(2004)}]{re:white04}
\bibinfo{author}{\bibfnamefont{S.~R.} \bibnamefont{White}} \bibnamefont{and}
  \bibinfo{author}{\bibfnamefont{A.~E.} \bibnamefont{Feiguin}},
  \bibinfo{journal}{Phys. Rev. Lett.} \textbf{\bibinfo{volume}{93}},
  \bibinfo{pages}{076401} (\bibinfo{year}{2004}).

\bibitem[{\citenamefont{Krylov}(1931)}]{re:krylov31}
\bibinfo{author}{\bibfnamefont{A.}~\bibnamefont{Krylov}},
  \bibinfo{journal}{Izvestija AN SSSR, Otdel. mat. i estest. nauk}
  \textbf{\bibinfo{volume}{VII}}, \bibinfo{pages}{491} (\bibinfo{year}{1931}).

\bibitem[{\citenamefont{{B. Bauer et al. (ALPS
  collaboration)}}(2011)}]{re:bauer11}
\bibinfo{author}{\bibnamefont{{B. Bauer et al. (ALPS collaboration)}}}
  (\bibinfo{year}{2011}), \bibinfo{note}{arXiv:1101.2646},
  \urlprefix\url{http://arxiv.org/abs/1101.2646}.

\bibitem[{\citenamefont{Lanczos}(1950)}]{re:lanczos50}
\bibinfo{author}{\bibfnamefont{C.}~\bibnamefont{Lanczos}}, \bibinfo{journal}{J.
  Res. Nat. Bur. Stand.} \textbf{\bibinfo{volume}{45}}, \bibinfo{pages}{255}
  (\bibinfo{year}{1950}).

\bibitem[{\citenamefont{Dagotto}(1994)}]{re:dagotto94}
\bibinfo{author}{\bibfnamefont{E.}~\bibnamefont{Dagotto}},
  \bibinfo{journal}{Review of Modern Physics} \textbf{\bibinfo{volume}{66}},
  \bibinfo{pages}{763} (\bibinfo{year}{1994}).

\bibitem[{\citenamefont{Hochbruck and Lubich}(1999)}]{re:hochbruck99}
\bibinfo{author}{\bibfnamefont{M.}~\bibnamefont{Hochbruck}} \bibnamefont{and}
  \bibinfo{author}{\bibfnamefont{C.}~\bibnamefont{Lubich}},
  \bibinfo{journal}{BIT} \textbf{\bibinfo{volume}{39}}, \bibinfo{pages}{620}
  (\bibinfo{year}{1999}).

\bibitem[{\citenamefont{Hochbruck and Lubich}(1997)}]{re:hochbruck97}
\bibinfo{author}{\bibfnamefont{M.}~\bibnamefont{Hochbruck}} \bibnamefont{and}
  \bibinfo{author}{\bibfnamefont{C.}~\bibnamefont{Lubich}},
  \bibinfo{journal}{SIAM J. Numer. Anal.} \textbf{\bibinfo{volume}{34}},
  \bibinfo{pages}{1911} (\bibinfo{year}{1997}).

\bibitem[{\citenamefont{Saad}(2003)}]{re:saad03}
\bibinfo{editor}{\bibfnamefont{Y.}~\bibnamefont{Saad}}, ed.,
  \emph{\bibinfo{title}{Iterative Methods for Sparse Linear Systems}}
  (\bibinfo{publisher}{Pws Pub Co}, \bibinfo{address}{Philadelphia},
  \bibinfo{year}{2003}).

\bibitem[{\citenamefont{White}(1996)}]{re:white96}
\bibinfo{author}{\bibfnamefont{S.}~\bibnamefont{White}},
  \bibinfo{journal}{Phys. Rev. Lett.} \textbf{\bibinfo{volume}{77}},
  \bibinfo{pages}{3633} (\bibinfo{year}{1996}).

\bibitem[{\citenamefont{Alvarez}(2009)}]{re:alvarez09}
\bibinfo{author}{\bibfnamefont{G.}~\bibnamefont{Alvarez}},
  \bibinfo{journal}{Computer Physics Communications}
  \textbf{\bibinfo{volume}{180}}, \bibinfo{pages}{1572} (\bibinfo{year}{2009}).

\bibitem[{\citenamefont{Alvarez}(2010)}]{re:alvarez11}
\bibinfo{author}{\bibfnamefont{G.}~\bibnamefont{Alvarez}}
  (\bibinfo{year}{2010}), \bibinfo{note}{arXiv:1003.1919}.

\bibitem[{\citenamefont{Knuth}(1992)}]{re:knuth92}
\bibinfo{author}{\bibfnamefont{D.~E.} \bibnamefont{Knuth}},
  \emph{\bibinfo{title}{Literate Programming}} (\bibinfo{publisher}{Center for
  the Study of Language and Information}, \bibinfo{address}{Stanford,
  California}, \bibinfo{year}{1992}).

\bibitem[{\citenamefont{Slater and Koster}(1954)}]{re:slater54}
\bibinfo{author}{\bibfnamefont{J.~C.} \bibnamefont{Slater}} \bibnamefont{and}
  \bibinfo{author}{\bibfnamefont{G.~F.} \bibnamefont{Koster}},
  \bibinfo{journal}{Phys. Rev.} \textbf{\bibinfo{volume}{94}},
  \bibinfo{pages}{1498} (\bibinfo{year}{1954}).

\bibitem[{\citenamefont{Noack et~al.}(1994)\citenamefont{Noack, White, and
  Scalapino}}]{re:landau94}
\bibinfo{author}{\bibfnamefont{R.}~\bibnamefont{Noack}},
  \bibinfo{author}{\bibfnamefont{S.}~\bibnamefont{White}}, \bibnamefont{and}
  \bibinfo{author}{\bibfnamefont{D.}~\bibnamefont{Scalapino}}, in
  \emph{\bibinfo{booktitle}{Computer Simulations in Condensed Matter Physics
  VII}}, edited by \bibinfo{editor}{\bibfnamefont{D.}~\bibnamefont{Landau}},
  \bibinfo{editor}{\bibfnamefont{K.}~\bibnamefont{Mon}}, \bibnamefont{and}
  \bibinfo{editor}{\bibfnamefont{H.}~\bibnamefont{Sch{\"u}ttler}}
  (\bibinfo{publisher}{Spinger Verlag}, \bibinfo{address}{Heidelberg, Berlin},
  \bibinfo{year}{1994}).

\bibitem[{\citenamefont{Al-Hassanieh et~al.}(2008)\citenamefont{Al-Hassanieh,
  Reboredo, Feiguin, Gonzalez, and Dagotto}}]{re:al-hassanieh08}
\bibinfo{author}{\bibfnamefont{K.~A.} \bibnamefont{Al-Hassanieh}},
  \bibinfo{author}{\bibfnamefont{F.~A.} \bibnamefont{Reboredo}},
  \bibinfo{author}{\bibfnamefont{A.~E.} \bibnamefont{Feiguin}},
  \bibinfo{author}{\bibfnamefont{I.}~\bibnamefont{Gonzalez}}, \bibnamefont{and}
  \bibinfo{author}{\bibfnamefont{E.}~\bibnamefont{Dagotto}},
  \bibinfo{journal}{Phys. Rev. Lett.} \textbf{\bibinfo{volume}{100}},
  \bibinfo{pages}{166403} (\bibinfo{year}{2008}).

\bibitem[{\citenamefont{Hubbard}(1963)}]{re:hubbard63}
\bibinfo{author}{\bibfnamefont{J.}~\bibnamefont{Hubbard}},
  \bibinfo{journal}{Proc. R. Soc. London Ser. A}
  \textbf{\bibinfo{volume}{276}}, \bibinfo{pages}{238} (\bibinfo{year}{1963}).

\bibitem[{\citenamefont{Hubbard}(1964{\natexlab{a}})}]{re:hubbard64a}
\bibinfo{author}{\bibfnamefont{J.}~\bibnamefont{Hubbard}},
  \bibinfo{journal}{Proc. R. Soc. London Ser. A}
  \textbf{\bibinfo{volume}{281}}, \bibinfo{pages}{401}
  (\bibinfo{year}{1964}{\natexlab{a}}).

\bibitem[{\citenamefont{Hubbard}(1964{\natexlab{b}})}]{re:hubbard64b}
\bibinfo{author}{\bibfnamefont{J.}~\bibnamefont{Hubbard}},
  \bibinfo{journal}{Proc. R. Soc. London Ser. A}
  \textbf{\bibinfo{volume}{281}}, \bibinfo{pages}{401}
  (\bibinfo{year}{1964}{\natexlab{b}}).

\bibitem[{\citenamefont{da~Silva et~al.}(2010)\citenamefont{da~Silva,
  Al-Hassanieh, Feiguin, Reboredo, and Dagotto}}]{re:diasdasilva10}
\bibinfo{author}{\bibfnamefont{L.~G. G. V.~D.} \bibnamefont{da~Silva}},
  \bibinfo{author}{\bibfnamefont{K.~A.} \bibnamefont{Al-Hassanieh}},
  \bibinfo{author}{\bibfnamefont{A.~E.} \bibnamefont{Feiguin}},
  \bibinfo{author}{\bibfnamefont{F.~A.} \bibnamefont{Reboredo}},
  \bibnamefont{and} \bibinfo{author}{\bibfnamefont{E.}~\bibnamefont{Dagotto}},
  \bibinfo{journal}{Phys. Rev. B} \textbf{\bibinfo{volume}{81}},
  \bibinfo{pages}{125113} (\bibinfo{year}{2010}).

\bibitem[{\citenamefont{Hager et~al.}(2004)\citenamefont{Hager, Jeckelmann,
  Fehske, and Wellein}}]{re:hager04}
\bibinfo{author}{\bibfnamefont{G.}~\bibnamefont{Hager}},
  \bibinfo{author}{\bibfnamefont{E.}~\bibnamefont{Jeckelmann}},
  \bibinfo{author}{\bibfnamefont{H.}~\bibnamefont{Fehske}}, \bibnamefont{and}
  \bibinfo{author}{\bibfnamefont{G.}~\bibnamefont{Wellein}},
  \bibinfo{journal}{J. Comp. Phys.} \textbf{\bibinfo{volume}{194}},
  \bibinfo{pages}{795} (\bibinfo{year}{2004}).

\bibitem[{\citenamefont{Chan}(2004)}]{re:chan04}
\bibinfo{author}{\bibfnamefont{G.~K.-L.} \bibnamefont{Chan}},
  \bibinfo{journal}{J. Chem. Phys.} \textbf{\bibinfo{volume}{120}},
  \bibinfo{pages}{3172} (\bibinfo{year}{2004}).

\bibitem[{\citenamefont{Kurashige and Yanai}(2009)}]{re:kurashige09}
\bibinfo{author}{\bibfnamefont{Y.}~\bibnamefont{Kurashige}} \bibnamefont{and}
  \bibinfo{author}{\bibfnamefont{T.}~\bibnamefont{Yanai}}, \bibinfo{journal}{J.
  Chem. Phys.} \textbf{\bibinfo{volume}{130}}, \bibinfo{pages}{234114}
  (\bibinfo{year}{2009}).

\bibitem[{\citenamefont{Yamada et~al.}(2009)\citenamefont{Yamada, Okumura, and
  Machida}}]{re:yamada09}
\bibinfo{author}{\bibfnamefont{S.}~\bibnamefont{Yamada}},
  \bibinfo{author}{\bibfnamefont{M.}~\bibnamefont{Okumura}}, \bibnamefont{and}
  \bibinfo{author}{\bibfnamefont{M.}~\bibnamefont{Machida}},
  \bibinfo{journal}{J. Phys. Soc. Jpn.} \textbf{\bibinfo{volume}{78}},
  \bibinfo{pages}{094004} (\bibinfo{year}{2009}).

\bibitem[{\citenamefont{Rinc{\'o}n et~al.}(2010)\citenamefont{Rinc{\'o}n,
  Garc{\'\i}a, and Hallberg}}]{re:rincon10}
\bibinfo{author}{\bibfnamefont{J.}~\bibnamefont{Rinc{\'o}n}},
  \bibinfo{author}{\bibfnamefont{D.~J.} \bibnamefont{Garc{\'\i}a}},
  \bibnamefont{and} \bibinfo{author}{\bibfnamefont{K.}~\bibnamefont{Hallberg}},
  \bibinfo{journal}{Comp. Phys. Comm.} \textbf{\bibinfo{volume}{181}},
  \bibinfo{pages}{1346} (\bibinfo{year}{2010}).

\bibitem[{\citenamefont{Amdahl}(1967)}]{re:amdahl67}
\bibinfo{author}{\bibfnamefont{G.}~\bibnamefont{Amdahl}}, in
  \emph{\bibinfo{booktitle}{AFIPS Conference Proceedings}}
  (\bibinfo{year}{1967}), vol.~\bibinfo{volume}{30}, p.
  \bibinfo{pages}{483–485},
  \urlprefix\url{http://www-inst.eecs.berkeley.edu/~n252/paper/Amdahl.pdf}.

\bibitem[{\citenamefont{Anderson et~al.}(1999)\citenamefont{Anderson, Bai,
  Bischof, Blackford, Demmel, Dongarra, Du~Croz, Greenbaum, Hammarling,
  McKenney et~al.}}]{laug}
\bibinfo{author}{\bibfnamefont{E.}~\bibnamefont{Anderson}},
  \bibinfo{author}{\bibfnamefont{Z.}~\bibnamefont{Bai}},
  \bibinfo{author}{\bibfnamefont{C.}~\bibnamefont{Bischof}},
  \bibinfo{author}{\bibfnamefont{S.}~\bibnamefont{Blackford}},
  \bibinfo{author}{\bibfnamefont{J.}~\bibnamefont{Demmel}},
  \bibinfo{author}{\bibfnamefont{J.}~\bibnamefont{Dongarra}},
  \bibinfo{author}{\bibfnamefont{J.}~\bibnamefont{Du~Croz}},
  \bibinfo{author}{\bibfnamefont{A.}~\bibnamefont{Greenbaum}},
  \bibinfo{author}{\bibfnamefont{S.}~\bibnamefont{Hammarling}},
  \bibinfo{author}{\bibfnamefont{A.}~\bibnamefont{McKenney}},
  \bibnamefont{et~al.}, \emph{\bibinfo{title}{{LAPACK} Users' Guide}}
  (\bibinfo{publisher}{Society for Industrial and Applied Mathematics},
  \bibinfo{address}{Philadelphia, PA}, \bibinfo{year}{1999}),
  \bibinfo{edition}{3rd} ed., ISBN \bibinfo{isbn}{0-89871-447-8 (paperback)}.

\end{thebibliography}

\end{document}